\documentclass[12pt,reqno]{article}
\usepackage{amsmath,hyperref,amssymb}
\usepackage{subfigure,graphicx,url}

\newcommand{\bea}{\begin{eqnarray}}
\newcommand{\eea}{\end{eqnarray}}
\newcommand{\be}{\begin{equation}}
\newcommand{\ee}{\end{equation}}

\allowdisplaybreaks \numberwithin{equation}{section}
\renewcommand{\Large}{\large} 

\DeclareSymbolFont{AMSa}{U}{msa}{m}{n}
\DeclareSymbolFont{AMSb}{U}{msb}{m}{n}
\DeclareMathSymbol{\fieldR}{\mathalpha}{AMSb}{"52}

\begin{document}
\begin{flushright} \small
SU-ITP-11/23\\
SLAC-PUB-14441
\end{flushright}
\bigskip
\begin{center}
 {\Large \bf
 Generalized Attractor Points in 
Gauged Supergravity  }\\[5mm]

 \

{\bf Shamit Kachru,$^{* \, \dag}$  Renata Kallosh,$^{*}$ and Marina Shmakova$^{\star}$}   \\[3mm]
 {\small\slshape
 * Department of Physics \emph{and} SITP, Stanford University \\
 Stanford, CA 94305-4060, USA \\
\medskip
 \dag Department of Particle Physics and Astrophysics, SLAC\\
 Menlo Park, CA 94309, USA
 \\
 \medskip
 $\star$ Kavli Institute for Particle Astrophysics and Cosmology,\\
 SLAC and Stanford University, Menlo Park, CA 94309, USA\\
\medskip
 {\upshape\ttfamily kallosh@stanford.edu, skachru@stanford.edu, shmakova@slac.stanford.edu}
\\[3mm]}
\end{center}

\

\vspace{5mm}  \centerline{\bfseries Abstract}
\medskip

The attractor mechanism governs the near-horizon geometry of extremal black holes in ungauged 4D N=2 supergravity theories and in Calabi-Yau compactifications of string theory.
In this paper, we study a natural generalization of this mechanism to solutions of arbitrary 4D N=2 gauged supergravities.
We define generalized attractor points as solutions of an ansatz which reduces the Einstein, gauge field, and scalar equations of motion to algebraic equations.
The simplest generalized attractor geometries are characterized by non-vanishing constant anholonomy coefficients in an orthonormal frame.
Basic examples include Lifshitz and Schr\"odinger solutions, as well as AdS and dS vacua. 
There is a generalized attractor potential whose critical points are the attractor points, and its extremization explains the algebraic nature of the
equations governing both supersymmetric and non-supersymmetric attractors.

\bigskip
\newpage

\tableofcontents

\newpage
\section{Introduction}

The study of the attractor mechanism has been a rich source of insights,  both into the physics of black holes \cite{Ferrara:1995ih,Ferrara:1996dd} and into more formal aspects of string theory \cite{Moore,OSV}.  In the best understood
case of BPS black holes in 4D N=2 (ungauged) supergravity, the original papers \cite{Ferrara:1995ih,Ferrara:1996dd} demonstrated that (vector multiplet) scalars are
attracted to universal fixed points at the black hole horizon, independent of their asymptotic values at infinity.
These fixed points are functions only of the charges of the black hole and the geometry of the moduli space, and can be found by minimizing a suitable attractor potential.
We will call the solutions where the scalars take the attractor values even asymptotically, so they remain constant instead of undergoing an attractor flow,
and where the gauge fields are also constants when expressed in terms of a suitable tetrad, attractor points.  In this language, the attractor points appearing
in the original papers
\cite{Ferrara:1995ih,Ferrara:1996dd} are the $AdS_2 \times S^2$ near-horizon geometries of the associated extremal black holes.

Several recent developments have encouraged us to revisit the attractor mechanism and the classification of attractor points in a more general context.  On the one hand,
there have been serious efforts to generalize the attractor mechanism to N=2 gauged supergravity solutions; see for example the recent work \cite{Hristov:2010eu,Cassani:2009na} and references therein.  We are able to extend
these results to provide a simple characterization of the attractor points for general N=2 gauged supergravities.

On the other hand, largely motivated by developments
at the interface of AdS/CFT and condensed matter physics \cite{adscmt}, there has been intense investigation of solutions with more general asymptotics than Minkowski
or AdS.
These more general asymptotic metrics can reflect, for instance, the symmetry groups of holographically dual field theories with Galilean \cite{Son,McGreevy} or Lifshitz \cite{KLM}
symmetries, which typically arise in condensed matter systems that do not enjoy emergent Lorentz invariance in the infrared.   We will find that the
emergence of such metrics in N=2 gauged supergravities is closely related to the attractor mechanism, and such metrics characterize generic classes of attractor
points with null or timelike Killing vectors.
In all of these generalized attractor solutions, the scalar fields take constant values, the gauge
fields and Riemann curvature are constants when expressed in terms of a suitable orthonormal tetrad, and the
equations of motion simplify from differential equations to algebraic equations.

We explain our strategy below and apply it to the case of D=4 N=2 gauged supergravity; similar ideas should of course work for further extended supergravities and other dimensions.
As in the past, one may expect that many features of the attractor points we find will be universal, and will be present in more general theories with and without
supersymmetry.

While to our knowledge the generality of circumstances in which one can find attractor points governed by algebraic equations is a novel result of our work,
several recent papers played an important role in motivating us to seek these simplifications.  The papers \cite{Cassani,Halmagyi:2011xh}, which instigated our work,
explored Lifshitz and Schr\"odinger solutions in particular gauged N=2 supergravities.   The earlier papers
\cite{Sch,10dim,Perlmutter} studied similar solutions from a higher-dimensional perspective; their low-energy effective theories are similar gauged supergravities.
The papers \cite{Goldstein} investigated the attractor mechanism for a
class of non-supersymmetric black branes, mostly focusing on attractor flows with AdS asymptotics and Lifshitz near-horizon geometries.  Our work is a step
towards extending the considerations of these papers to find the most general attractors in the context of fully general N=2 gauged supergravities, where (somewhat
surprisingly) the problem is tractable. We also find a geometric characterization of the simplest generalized attractors: their geometry has constant anholonomy coefficients,
which implies that the tangent space curvature components are constant and non-singular.

The organization of this paper is as follows.  In \S2, we define the generalized attractors in gauged supergravity and explain their geometry, in particular, the importance of the tangent space and the concept of anholonomy coefficents. In \S3, we introduce the generic class of 4D N=2 gauged supergravities we will study. We generalize the definition of `fermionic shifts' to the case of solutions with constant vector fields and fluxes. We  show that for attractor points, configurations with constant scalar fields and constant gauge fields in an appropriate sense, the result is a set of
algebraic equations which replace the Einstein equations and the other bosonic equations of motion. In \S4 we study supersymmetric attractors. We
outline a basic ingredient in our strategy, which is to use Killing spinor identities to help guarantee that certain
supersymmetric configurations satisfy all equations of motion.     In \S5 we describe a generalized attractor potential whose extrema determines
the attractor points.   \S6 contains a few simple examples of generalized attractor points, while we close with a discussion in \S7.
In Appendix A we present the set of all constant anholonomy coefficients specifying the anisotropic Lifshitz geometry and the set of all constant non-singular tangent space components of the curvature tensor, contrasted with the spacetime dependent curved space components of the curvature tensor. In Appendix B we present the analogous data for the Schr\"odinger geometry.

\section{Generalized Attractors in D=4 Supergravity}

We are interested in finding a generalization of the attractor mechanism of \cite{Ferrara:1995ih,Ferrara:1996dd} for cases which include not only constant values of the scalars $\phi^i$ and tangent space field strengths  $F_{ab}= e_{a}^{\mu} e_{b}^{\nu} F_{\mu\nu}$ (fluxes), but also constant values of the tangent space vectors $A_a= e^\mu_a A_\mu$.
While in the ungauged N=2 supergravity the attractor data depend only on the
field strengths $F_{ab}$, in the gauged theory the Lagrangian has additional dependence on $g\,  A_a$ (where $g$ is the gauge coupling) that affects the attractor values of the moduli; this explains our interest in solutions with constant $A_a$.  (Note that, in general, a constant field strength does not require constant vectors).

In the supergravity literature, the attractors are known as critical points of the supergravity action, gauged or ungauged, which solve an algebraic form of the equations of motion -- in particular, algebraic BPS conditions.  At the attractor points, scalars, fluxes and curvature are constant in the tangent space. Examples include the $AdS_2 \times S^2$ near-horizon geometries of  extremal black holes with constant scalars and constant values of $F_{01}$ and $F_{23}$.  Our generalized attractors in gauged supergravity have an important non-trivial new ingredient, which enters in the algebraic equations determining the attractor point: constant background vector fields.\footnote{It is important in the generic gauged supergravity case to have constant $A_a$, because unlike in the case of ungauged attractors, the gauge fields (and not just the field strengths) enter directly into the expression for the
attractor potential; see \S5.  Therefore, we will take this as part of the definition of a generalized attractor point.}

More precisely, the special solutions singled out by \cite{Ferrara:1995ih,Ferrara:1996dd} are the attractor points where the scalars and fluxes take constant values in the entire solution,
determined by the black hole charges and the geometry of the moduli space.  We will find that the most elegant generalization to fully generic gauged D=4 N=2
supergravities requires that we impose the following conditions: the scalars should be constant, and the gauge fields and field strengths should take constant values in the tangent space, in an orthonormal frame.

 A  simple new condition (which is not strictly necessary, as we shall see below, but which characterizes the simplest generalized attractors with frozen scalars, gauge fields, and field strengths) is that one should have { \it constant values of the anholonomy coefficients} $c_{ab}{}^c$ which are  sometimes called Ricci rotation coefficients.  See for example \cite{Ortin:2004ms} for details on supergravity tangent space and vierbein formalism, including the discussion of non-holonomic frames.
 The relation that follows between constant field strengths $F^{\Lambda}_{ab}$ and constant Abelian vectors $A^\Lambda_a$ is
\be
\label{gfe}
F^{\Lambda}_{ab} = c_{ab}^c A_c^{\Lambda}~.
\ee
The easiest way to solve this equation with constant gauge fields and field strengths, is if the anholonomy coefficients themselves are constant.  We will see in \S3 that, in fact, to get algebraic equations of motion with a fully generic ansatz for the gauge fields, constant anholonomy will be required.
 In this case, one can directly prove that the tangent space Riemann-Christoffel curvature is constant as well, and so in the simplest ansatz
 \be
 F_{ab}^\Lambda  = c_{ab}{}^c A_c^\Lambda=\rm const\  \qquad \Rightarrow \qquad R_{abc}{}^d=const
\label{genAttr} \ee
If in certain directions in the tangent space the fluxes vanish,
$F_{\hat a \hat b}=0$, or the vectors vanish, $A_{\hat c}=0$,
 then the restrictions on the corresponding components of the anholonomy coefficients can be relaxed.  It is interesting that spaces with all constant anholonomy coefficients have constant Riemann-Christoffel tangent space curvature  which is obviously non-singular. This property is very much in the spirit of the black hole attractors with regular horizon geometry, where scalars reach a particular charge dependent constant value, as opposed to black holes with a singular horizon geometry where scalars have runaway behaviour. The feature of all constant anholonomy coefficients guarantees non-singular constant Riemann-Christoffel curvature in the tangent space.\footnote{It is known that some of
 these attractor solutions do have large tidal forces at loci in the space-time; this is true, for instance, of the Lifshitz solutions we discuss in \S6.  We have nothing new to say about the proper interpretation of that here.}

 Because of the caveat that when certain fluxes or vectors vanish one can relax the corresponding constancy condition on the anholonomy coefficients, constant anholonomy
  is a sufficient but not a necessary requirement for attractors with constant curvature and fluxes.
  The earliest examples of attractors in fact exploit this caveat, and do not have constant anholonomy.
   In particular, in ungauged supergravity, the attractor condition requires only that the non-vanishing
\be
F_{ab}^\Lambda = e_{a}^{\mu} e_{b}^{\nu} (\partial_{\mu} A_\nu^\Lambda -\partial_{\nu} A_\mu^\Lambda)
\ee
 are  constant at the attractor point. For example in case of $AdS_2\times S^2$ attractors in \cite{Ferrara:1996dd} ,
 \be
 F= {1\over q} \, dr \wedge dt={1\over q} \,  e^0\wedge e^1
 \ee and
 \be
 G= p \sin \theta \, d\theta\wedge d\phi= p \, e^2\wedge e^3~.
 \ee
  The tangent space magnetic field in this solution is
  \be
  A_3= e_3^\phi A_\phi \sim {\cos \theta\over \sin \theta}
  \ee
 and the corresponding anholonomy coefficient $c_{23}{}^3$ is  not constant -- it depends  on $\theta$, so that the flux $G_{23}=p$ is constant.  This gives an example where the Riemann-Christoffel tangent space curvature $R_{abc}{}^d$  is constant and  non-singular, without fully constant anholonomy.   In addition, the gauge field is not constant.
 However, in the ungauged case, as we mentioned above, the gauge field (as opposed to the field strength) does not enter into the attractor potential, and one need not include constant $A_a$ as part of the definition of an attractor.
 Nevertheless,  since our interest is the generalization to generic gauged supergravity, we will for the most part focus on the simplest generalized attractors below.
 These solutions do exhibit constant anholonomy, in addition to constant
 vectors, field strengths, and Riemann curvature.  Within this class, we will find generic solutions which have emerged as spacetimes of independent interest in
 the search for gravity duals of non-relativistic field theories.

In what follows, we derive the basic equations governing generalized attractors in gauged supergravity.  Because the formalism of N=2 gauged supergravity can obscure
the essential points, we give an overview of where we are going in the next subsection.  We move on to do the more detailed supergravity analysis in \S3.

\subsection{Geometry of Supergravity Attractors}

Here we briefly summarize some features of the geometry of supergravity.  These include concepts like vierbeins, tangent space, orthonormal frame, spin connection, anholonomy coefficients, torsion and curvature tensors; see for example \cite{Ortin:2004ms} for further details.  We will derive here the following statements which are relevant for
generalized attractors:

i) in a space-time with constant anholonomy coefficients $c_{ab}^c$ (like our simplest generalized attractors), there is a constant
non-singular curvature tensor which is determined by the $c_{ab}^c$.

 ii) the generalized attractor assumption that there are both constant fluxes and constant gauge fields is most simply satisfied with constant anholonomy coefficients, and yields
 eq. (\ref{genAttr}).  More general solutions (as exemplified by the $AdS_2 \times S^2$ solutions above) can also exist, but constant anholonomy is the simplest case (and,
 as we will see in \S3, the case that guarantees algebraic equations of motion with a fully generic ansatz of constant tangent-space gauge fields and field strengths).

Given the metric $ds^2= g_{\mu\nu} (x) dx^\mu dx^\nu$,   the vierbein $e_\mu^a$ is related to the metric:
\be
g_{\mu\nu} (x) = e_\mu^a(x) e_\nu ^b (x)\eta_{ab} \qquad a=0,1,2,3
\ee
Here $\eta_{ab}$ is the  tangent space Minkowski metric.  In general relativity, the Lorentz covariant tangent space of supergravity is related to an orthonormal frame. Supergravity fermions  live in tangent space: the Dirac equation in curved space is $\gamma^a e^\mu _a D_\mu \psi= \gamma^\mu  D_\mu \psi=0$. Here $\gamma^a$ is the tangent space numerical Dirac $\gamma$-matrix,
$e^\mu_a(x) $ is an inverse vierbein, and $\gamma^\mu(x) = \gamma_a e_a^\mu (x)$.  These objects satisfy the relations:
\be
e^a_\mu e^\mu_b=\delta^a_b \qquad e^a_\mu e^{\mu b}=\eta^{a b}
\ee
It is convenient to define the vierbein form
\be
e^a\equiv e_\mu^a(x) dx^\mu \qquad ds^2 = \eta_{ab} e^a e^b
\ee
 and its dual
 \be
 \tilde e_a \equiv e_a^{\mu}\partial_{\mu}\equiv \partial_a~.
 \ee
 The anholonomy coefficients $c_{ab}^{~~c}$ are defined via the commutator of the dual vierbein forms
 $[ \tilde e_a^{\mu}(x) \partial_{\mu}, \tilde e_b^{\nu}(x) \partial_{\nu}]  \equiv [ \partial_a, \partial_b] \equiv  c_{ab}^{~~c}  e_c^{\lambda}\partial_{\lambda}$:
\be
[\tilde e_a, \tilde e_b] \equiv  c_{ab}^{~~c} \tilde e_c~, \qquad c_{ab}{}^c= e_a^{\mu} e_b^{\nu} ( \partial_\nu e_\mu^c -\partial_\mu e_\nu^c )
\ee
A Lorentz covariant derivative  in the tangent space of supergravity (acting on spinors and tangent space Abelian vectors\footnote{In non-Abelian cases, the non-Abelian connection term $e_a^\mu A_\mu^\Lambda {\cal T}_\Lambda$ is present, where ${\cal T}_\Lambda$ is the non-Abelian gauge symmetry generator.}) requires the spin connection $\omega$
\be
D_a= \partial_a + \omega_a{}^{bc} {\cal M}_{bc}= e_a^{\mu}( \partial_{\mu} + \omega_\mu{} ^{bc} {\cal M}_{bc})\, ,
\ee
where ${\cal M}_{ab}$ is the Lorentz generator. The commutator of two such derivatives, in general, defines the torsion and the curvature tensors of the tangent space:
\be
 [D_a, D_b] =  T_{ab}{}^c D_c+ R_{ab}{}^{ cd}  {\cal M}_{cd}
  \ee
where  the tangent space torsion tensor is
 \be
 T_{ab}{}^c= c_{ab}^{~~c}  + \omega_a{} ^c{}_b- \omega_b{} ^c{}_a
 \ee
and the tangent space curvature is
\be
R_{abc}{}^{ d}= \partial_a \omega_{bc}{}^d- \partial_b \omega_{ac}{}^d- \omega_{ac}{}^e \omega_{be}{}^ d
+ \omega_{bc}{}^e \omega_{ae}{}^ d - c_{ab}{}^e \omega_{ec}{}^d~.
\ee
In the case of a Riemannian spacetime, the torsion is absent and we have a relation between the anholonomy coefficients and the spin connection
 \be
 \label{torsion}
 T_{ab}{}^c= 0 \qquad \Rightarrow \qquad c_{ab}{}^c=  \omega_a{} ^c{}_b- \omega_b{} ^c{}_a \ee
  The inverse relation defines the spin connection in terms of the anholonomy coefficients
  \be
 \omega_{a, bc} = {1\over 2} (c_{a b, c} -c_{ac, b} -c_{bc, a})
\label{spincon} \ee
It follows that when the anholonomy coefficients are constants.
\be
\partial_a c_{bc}{}^d=0~,
\ee
  the spin connection and  the curvature are constant
\be
R_{abc}{}^{ d}= - \omega_{ac}{}^e \omega_{be}{}^ d
+ \omega_{bc}{}^e \omega_{ae}{}^ d - c_{ab}{}^e \omega_{ec}{}^d~.
\label{attrcurv}\ee
When we replace the constant spin connections  in (\ref{attrcurv}) by their expression in terms of  the anholonomy coefficients  in (\ref{spincon}), we find the non-singular curvature tensor as a function of constant anholonomy coefficients
\be
R_{abc}{}^{ d}= R_{abc}{}^{ d} [c_{ef}{}^g]
\ee
This proves claim i) in the beginning of this subsection.

 Our attractor point condition requires vectors and fluxes to be constant. This is possible only if
 \be
 A_a^\Lambda= e^\mu_a A_\mu^\Lambda={\rm const} \qquad F_{ab}^\Lambda= e_{a}^{\mu} e_{b}^{\nu} F_{\mu\nu}^\lambda={\rm const}
 \ee
 since
 \be
 F_{ab}= e_a^\mu e_b^\nu( \partial _{\mu} e_{\nu}^c -\partial _{\nu} e_{\mu}^c) A_c = c_{ab}{}^c A_c~.
 \ee
Thus, if  both $A_c$ and $ F_{ab}$ are constant, consistency is most obvious if the anholonomy coefficients $c_{ab}^{~~c}$ are constant (in the cases that they relate
non-vanishing components of $A$ and $F$; as described in \S2.1, there are interesting special examples where some components of the gauge fields vanish, and one does
not require constant anholonomy in those directions).\footnote{Of course one may also contemplate non-trivial anholonomies with cancellations between different terms in the
sum over $c$, but the constant anholonomy attractors are the most elegant.  They are also the ones which solve algebraic equations of motion for a generic ansatz of constant tangent space fields.}

This justifies our claim ii) in the beginning of this section.


 We therefore define the generalized attractors of D=4 N $\geq$ 2 gauged or ungauged supergravity as
solutions of  equations of motion which are reduced to purely algebraic equations, where the tangent space curvature tensor, scalars, and fluxes are  constant. In the particular case of gauged supergravity, when also tangent space vectors are non-vanishing constants and the anholonomy coefficients are constant, we find an interesting set of attractors which we will study below.
Such solutions exist both with and without unbroken supersymmetry.

 To summarize this section: we have motivated the study of generalized attractors of
N=2, D=4 supergravity with the following features:
\be\boxed{
z^i= {\rm const}\, , \quad q^u={\rm const} \, ,  \quad  A_a={\rm const}\, ,   \quad F_{ab}=  {\rm const} \, , \quad c_{ab}{}^c={\rm const}\, , \quad R_{ab}{}^{ cd}= {\rm const}}
\label{attractor} \ee
where $z^i$ are scalars from the vector multiplets and $q^u$ are the quaternions from the hypermultiplets.
In the generic case where there are non-vanishing vectors at the attractor points, the attractor values of the scalars will depend on $A_a$ and $ F_{ab}$, and on all
of the parameters entering in the gauged supergravity `fermionic shifts' (which we define and study in \S3), which include the gauge couplings, the
Killing vectors gauging the isometries of the special K\"ahler and quaternionic K\"ahler vector multiplet and hypermultiplet moduli spaces, and the momentum maps.
We note here that our phrasing of the conditions for an attractor point in (\ref{attractor}) is not gauge invariant; a gauge transformation can clearly turn
constant scalars and vectors into non-constant scalars and vectors.  However, it is easy to check that the equations of motion transform in a gauge covariant
way, so having found a solution of the type (\ref{attractor}), a gauge transformation will preserve the fact that it solves the equations of motion, while 
making the solution look more complicated.  In \S3.2, we will further motivate choosing the condition of constant scalars (instead of e.g. covariantly constant scalars) 
in this definition of generalized attractors.

\section{Gauged N=2 D=4 Supergravity}

N=2, D=4 supergravities are of particular interest for several reasons.  On the one hand, the constraints which follow from N=2 supersymmetry make
such theories, to some extent, solvable.  On the other hand, they still exhibit a rich structure of quantum corrections and qualitatively new phenomena compared
to their counterparts with further extended supersymmetry.  Finally, as the low-energy effective theories arising from Calabi-Yau compactifications of type II
string theories, they enjoy a special role in studies of string compactifications and dualities as well.

The complete Lagrangian for gauged N = 2
supergravity \cite{de Wit} in presence of $n_V$ Abelian vector multiplets and
$n_H$ hypermultiplets, with generic gauging of the scalar manifold isometries, is presented in
\cite{Andrianopoli:1996cm}. It was used in recent applications (searching for solutions with unbroken supersymmetry) in \cite{Hristov:2010eu,Cassani:2009na,Cassani,Halmagyi:2011xh}.  Here, we
will use various features of generic  BPS solutions that were derived in  \cite{Hristov:2010eu}, and apply them to the specific case of attractor points with frozen scalars.
We follow the notation of \cite{Hristov:2010eu}.

Gauging modifies the Lagrangian in several ways: it requires the use of gauge-covariant  derivatives for the
 scalars, and also the addition of  a scalar potential. The bosonic Lagrangian is
\begin{align}\label{lagr}
\begin{split}
\mathcal L&=-\frac{1}{2}R(g)+g_{i\bar
\jmath}\nabla^\mu z^i
\nabla _\mu
{\bar z}^{\bar \jmath} + h_{uv}\nabla^\mu q^u \nabla_\mu q^v\\&+
I_{\Lambda\Sigma}F_{\mu\nu}^{\Lambda}F^{\Sigma\,\mu\nu}
+\frac{1}{2}R_{\Lambda\Sigma}\epsilon^{\mu\nu\rho\sigma}
F_{\mu\nu}^{\Lambda}F^{\Sigma}_{\rho\sigma} - g^2V (z, \bar{z},
q)\ .
\end{split}
\end{align}
Here $i=1,...,n_V$, $\Lambda = 0,...,n_V$ and $u = 1,..., 4 n_H$.
The covariant derivatives on scalars are
\be
\label{cov}
\nabla_{\mu} z^i \equiv
\partial_{\mu} z^i +g { k}^i_{\Lambda} A^{\Lambda}_{\mu}
\, , \qquad \nabla_{\mu} q^u \equiv
\partial_{\mu} q^u +g { k}^u_{\Lambda} A^{\Lambda}_{\mu}
\ee
and they define the gauging of isometries of the vector and hypermultiplet
scalar  manifolds with Killing vectors  ${
k}^i_{\Lambda}(z)$ and ${
k}^u_{\Lambda}(q)$, respectively,  and coupling constant $g$.
The functions $I_{\Lambda\Sigma}$ and $R_{\Lambda\Sigma}$ are determined by special geometry in terms of
the holomorphic prepotential, or in more general cases, in terms of the periods $X^\Lambda$ and the dual periods $F_{\Lambda}$; we will not
need the explicit formulae here.

The scalar potential depends on  the Killing
vectors and the corresponding triplet of quaternionic moment maps
$P^x_{\Lambda}$:
\begin{equation}\label{pot2}
V(z, \bar{z},
q) =\Big (g_{i\bar
\jmath} {k}^i_\Lambda {k}^{\bar
\jmath}_\Sigma   +  4 h_{uv} {k}^u_\Lambda {k}^v_\Sigma \Big ) {\bar L}^\Lambda
L^\Sigma + (g^{i\bar \jmath}f_i^\Lambda {\bar f}_{\bar
\jmath}^\Sigma -3{\bar L}^\Lambda L^\Sigma)P^x_\Lambda
P^x_{\Sigma}\ ,
\end{equation}
where
\begin{equation}
L^\Lambda={\rm e}^{{\cal K}(z, \bar z)/2}X^\Lambda(z) \ ,\qquad f_i^\Lambda= {\rm e}^{{\cal K}/2}D_iX^\Lambda\
\end{equation}
and $X^\Lambda$ are the holomorphic sections governing the special K\"ahler geometry.
The action is invariant under the following
supersymmetry variations (up to higher order terms in fermions):
\begin{align}
\delta_\varepsilon\lambda^{iA}&=i\nabla_\mu z^i
\gamma^\mu\varepsilon^A + G_{\mu\nu}^{-i}
\gamma^{\mu\nu}\epsilon^{AB}\varepsilon_B+i gg^{i\bar \jmath}{\bar
f}^\Lambda_{\bar \jmath}P^x_\Lambda\sigma_x^{AB}\varepsilon_B\
,\label{susygluino}
\\
\delta_\varepsilon \zeta_\alpha &= i\,
\mathcal{U}^{B\beta}_u\nabla_\mu q^u \gamma^\mu \varepsilon^A
\epsilon_{AB}\mathbb{C}_{\alpha\beta} +
2g\,\mathcal{U}^A_{\alpha\,u}{\tilde k}^u_\Lambda {\bar L}^\Lambda\
\varepsilon_A \ ,\label{susy-hyperino}
\\
\delta_\varepsilon \psi_{\mu A}&=\nabla_\mu\varepsilon_A +
T^-_{\mu\nu}\gamma^\nu \epsilon_{AB}\varepsilon^B + ig S_{A B}
\gamma_\mu\varepsilon^B\ ,\label{susy-gravi}
\end{align}
where $\lambda^{iA}, \zeta_\alpha$ and $\psi_{\mu A}$ are the gauginos, hyperinos and gravitinos respectively, and $A=1,2$. The supercovariant derivative in \eqref{susy-gravi} is defined as
\begin{align}\label{eq:def-covariant-epsilon}
  \nabla_\mu \varepsilon_A = \left(\partial_\mu - \frac 14
    \omega_\mu^{ab} \gamma_{ab}\right) \varepsilon_A + \frac i 2 {\cal A}_\mu
  \varepsilon_A + \omega_{\mu A}{}^B \varepsilon_B\ .
\end{align}
The connections ${\cal A}_\mu$ and $\omega_{\mu A}{}^B$ are associated to
the special K\"ahler and quaternionic K\"ahler manifolds,
respectively; see \cite{Andrianopoli:1996cm} for more details.
Here the gravitino field strength and mass matrix are
\begin{equation}\label{mass-gravitino}
T^-_{\mu\nu}\equiv 2iF^{\Lambda\,-}_{\mu\nu}\,I_{\Lambda\Sigma} L^\Sigma \ ,\qquad
S_{AB}\equiv\frac{i}{2}(\sigma_x)_{AB}P^x_\Lambda L^\Lambda\ ,
\end{equation}
and
$
F_{\mu\nu}^{\Lambda\,-}=i{\bar L}^\Lambda
T^-_{\mu\nu}+2f_i^\Lambda G^{i\,-}_{\mu\nu}\
$ (where the $^-$ denotes the anti-selfdual part of a two-index tensor).

\subsection{Fermionic shifts in generalized attractors of gauged supergravity}
In the supergravity literature, `fermionic shifts' were traditionally associated with the part of the supersymmetry transformations of fermions which is present in vacua with constant scalars. Here we slightly generalize the definition of fermionic shifts, to include the presence of constant (tangent space) vectors and fluxes. Thus we drop from (\ref{susygluino})-(\ref{susy-hyperino}) terms with derivative of scalars and Lorentz covariant derivative on spinors
\be
\partial_\mu z^i=0\, , \qquad \partial_\mu q^u=0\, ,
\label{frozen}\ee
but keep $A_a$ and $F_{ab}$. Since these are constant at attractors and always contracted with numerical $\gamma$-matrices in the tangent space, these extra terms in the fermionic shifts are also constants and may be treated on the same footing as terms depending on constant scalars.
The remaining terms in (\ref{susygluino})-(\ref{susy-hyperino}) with account of (\ref{frozen}) define the fermionic shifts in gauged supergravity attractors.

For the gluinos we have the shift $ \tilde \delta^B   \lambda ^{i A}$ defined as
\begin{align}
 \tilde \delta  \lambda ^{i A} \Rightarrow  (\tilde \delta^B  \lambda ^{i A})  \varepsilon_B=ik^i_\Lambda A_a^\Lambda
\gamma^a\varepsilon^A + G_{ab}^{-i}
\gamma^{ab}\epsilon^{AB}\varepsilon_B+i gg^{i\bar \jmath}{\bar
f}^\Lambda_{\bar \jmath}P^x_\Lambda\sigma_x^{AB}\varepsilon_B\
.\label{gluinoshift}
\end{align}
For hyperinos we have the shift $ \tilde \delta^B   \zeta_\alpha$ defined as
\begin{align}
\tilde \delta \zeta_\alpha\Rightarrow  (\tilde \delta^A  \lambda ^{i A})  \varepsilon_A  =
i \mathcal{U}^{B\beta}_u k^u_\Lambda A_a^\Lambda   \gamma^a \varepsilon^A
\epsilon_{AB}\mathbb{C}_{\alpha\beta} +
2g\,\mathcal{U}^A_{\alpha\,u}{ k}^u_\Lambda {\bar L}^\Lambda\
\varepsilon_A \ .\label{susy-hyperino1}
\end{align}
Finally, for the gravitino in the tangent space, $\psi_{a A}= \psi_{\mu A} e^\mu_a$, we are interested in an integrability condition for the unbroken supersymmetry condition $\delta_\varepsilon \psi_{a A}=0$, which we present in the form
\be
\delta_\varepsilon \psi_{a A}=D_a \varepsilon_A + (\tilde \delta^B \psi_{a A})\varepsilon_B~.
\, ,\label{susy-gravi1}
\ee
Then the fermionic shift  $(\tilde \delta^B \psi_{a A})$ is
 \be
 \tilde \delta\psi_{a A} = (\tilde \delta^B \psi_{a A}) \varepsilon_B=
 \frac i 2 {\cal A}_a
  \varepsilon_A + \omega_{a A}{}^B \varepsilon_B +
T^-_{ab}\gamma^b \epsilon_{AB}\varepsilon^B + ig S_{A B}
\gamma_a \varepsilon^B ~.
\label{susy-gravitino} \ee
 The connections ${\cal A}_\mu= e_\mu^a {\cal A}_a$ and $\omega_{\mu A}{}^B= e_\mu^a \omega_{\mu A}{}^B$ are associated to
the special K\"ahler and quaternionic K\"ahler manifolds
\cite{de Wit,Andrianopoli:1996cm,Hristov:2010eu} and
 \be
 D_a \varepsilon_A= \left(\partial_\mu - \frac 14
    \omega_\mu^{ab} \gamma_{ab}\right) \varepsilon_A
 \ee

 \subsection{Algebraic bosonic  equations of motion at the attractor points}

{\it Vector equations}

 The dual field strength $G_{\Lambda\mu\nu}$ is defined as a derivative of the action with respect to
$F^{\Lambda}_{\mu \nu}$. In absence of gauging
\begin{align}\label{eq:defg}
G_{\Lambda}{}_{\mu \nu} \equiv R_{\Lambda
    \Sigma} F^{\Sigma}_{\mu \nu} - \frac 12 I_{\Lambda \Sigma}\,
  \epsilon_{\mu \nu \gamma \delta} F^{\Sigma \gamma \delta}\ .
\end{align}
The Maxwell equations and Bianchi identities take a
simple form
\begin{equation}\label{eq:blsmaxwell}
    \epsilon^{\mu \nu \rho \sigma} \partial_{\nu} G_{\Lambda}{}_{\rho \sigma} =
    0, \quad \epsilon^{\mu \nu \rho \sigma} \partial_{\nu} F^{\Lambda}_{\rho \sigma} =
    0\ ,
\end{equation}
and $(F^\Lambda, G_\Lambda)$ transforms as a vector under electric-magnetic duality transformations.
Gauging breaks the duality symmetry: the Bianchi identities $\epsilon^{\mu \nu \rho \sigma} \partial_{\nu} F^{\Lambda}_{\rho \sigma} =
    0
$ remain the same, however, the gauge field equations of motion change since the action now has additional dependence on $A_\mu^\Lambda$, not only on $F^{\Lambda}_{\mu \nu}$:
\begin{equation}\label{maxwellt}
\epsilon^{\mu \nu \rho \sigma} \partial_{\nu} G_{\Lambda\rho\sigma} =
- g( h_{u v} { k}^u_{\Lambda} \nabla^{\mu} q ^v+ g_{i\bar
\jmath} {1\over 2}({k}^i_\Lambda \nabla^{\mu}
  z^{\bar
\jmath} +hc)\ .
\end{equation}
We see now in (\ref{maxwellt}) that requiring covariantly constant scalars, instead of constant scalars, would be equivalent to causing the right hand side of 
this equation to vanish (and hence constitute some loss of generality).  Since some of the most interesting solutions which characterize holographic RG fixed points (like Lifshitz and Schr\"odinger solutions) 
arise with scalars which are constant (in an appropriate gauge) but not covariantly constant, we have chosen the former condition; we would like to include such
basic solutions in our definition of generalized attractors.  Of course, there may well
also be interesting solutions with covariantly constant scalars which are however not constant in any gauge; they do not fall under our current definition of
generalized attractors.

At the attractor points (\ref{attractor}) when the gauge fields are non-vanishing, these equations simplify:
\begin{equation}\label{maxwell}
\epsilon^{\mu \nu \rho \sigma} \partial_{\nu} G_{\Lambda\rho\sigma} =
- g( h_{u v} { k}^u_{\Lambda} A^{\mu\Sigma} k_\Sigma  ^v+ g_{i\bar
\jmath} {k}^i_\Lambda A^{\mu \Sigma}
  \bar k^{\bar
\jmath}_\Sigma) \ .
\end{equation}
They can also be given in the form
\begin{equation}\label{maxwell}
\epsilon^{\mu \nu \rho \sigma} \partial_{\nu} G_{\Lambda\rho\sigma} =
 A^{\mu\Sigma} C_{ \Sigma \Lambda}\ ,
\end{equation}
where $C_{ \Sigma \Lambda}$ is constant at the attractor point. In the tangent space we find
\be
\epsilon^ {abcd} D_b G_{\Lambda cd}=  A^{a\Sigma} C_{ \Sigma \Lambda}~.
\ee
At the attractor point, $G_{\Lambda cd}$ is constant, related to a constant $F_{ab}$ as follows:
\begin{align}\label{eq:defg}
G_{\Lambda}{}_{ab} \equiv R_{\Lambda
    \Sigma} F^{\Sigma}_{ab} - \frac 12 I_{\Lambda \Sigma}\
  \epsilon_{abcd} F^{\Sigma cd}\ .
\end{align}
This means that $\partial_b G_{\Lambda cd}$ drops out of the gauge field equation and we are left with an algebraic equation
\be
\partial_b G_{\Lambda cd}=0\,  \qquad \Rightarrow \qquad \epsilon^ {abcd} (D_b-e_b) G_{\Lambda cd}=  A^{a\Sigma} C_{ \Sigma \Lambda}~.
\ee
In the case of Abelian gauging, the connection in $(D_b-\partial_b)$ is just a constant spin connection $\omega_a{}^{b c}$
\be
\label{canh}
\epsilon^ {abcd} \omega_a{}^e{}_c G_{\Lambda ed} =  A^{a\Sigma} C_{ \Sigma \Lambda}~.
\ee
The fact that the gauge field equations of motion are algebraic at the attractor point is based on the generic properties of the attractor points defined above, and it is universal.

Notice that above, we assumed that the spin connection is constant, and with a fully generic (constant in tangent space) ansatz for the gauge fields and field strengths, the equation (\ref{canh}) is only obviously algebraic in this case.  But constancy of the spin connection implies, via (\ref{torsion}), that the anholonomy coefficients
$c_{ab}^{~~c}$ are constant as well.  This justifies our focus on the case of constant anholonomy, despite the fact that famous examples like $AdS_2 \times S^2$
satisfy algebraic equations with slightly different assumptions.  It might be interesting to explore a range of other ansatzes for the gauge fields and spin connections which
could result in algebraic equations as well.

{\it Einstein equations}

In a background with constant scalars, anholonomy, and curvature that also solves the vector field equations above, one can check that the Einstein equations
also reduce to algebraic equations.  Namely, in

\be
\label{eeq}
R_{ab}-{1\over 2} \eta_{ab} R = T_{ab}^{\rm attr}~,
\ee
direct inspection of the supergravity action shows that $T_{ab}^{\rm attr}$ depends only on the space-time independent solutions for the vectors and scalars, and therefore is constant. The left hand side is constant by the definition of the attractor point. Solving the Einstein equations simply requires an identification of constants which makes
(\ref{eeq}) it consistent.

{\it Scalar equations}

Assuming the vector and Einstein equations can be solved for some constant value of the scalars, the equations of motion for the scalars themselves reduce to
the condition that a suitable attractor potential (to be discussed in more detail in \S5) be extremized:
\be
\partial_{z^i} {\cal V}_{\rm attractor}= \partial_{q^u} {\cal V}_{\rm attractor}=0~.
\ee
Therefore, at extrema of ${\cal V}_{\rm attractor}$, we find a generalized attractor point.

So, we have demonstrated that finding generalized attractor points in gauged N=2 supergravity with
$
z^i,  q^u,   A_a^\Lambda, F_{ab}^\Lambda,  c_{ab}{}^c,  R_{ab}{}^{ cd}$ all constant, just requires one to solve {\it algebraic} equations of motion.
However, in general, this is not easy, and unbroken supersymmetry  helps.  This is the topic of \S4.

\section{Killing Spinors, BPS conditions and  Equations of Motion }

In supergravity, a useful shortcut to find particular  solutions of the equations of motion, which are second order differential equations, is to solve the conditions for unbroken supersymmetry.  For a classical background with vanishing Fermi fields $\psi$, this means one should impose
\be
\label{unbroken}
\delta \psi=0
\ee
where $\delta$ is the SUSY variation.
These are first order differential equations. They are called BPS conditions because solutions of these equations saturate a BPS bound.
Solving these first order differential equations is, in many cases, sufficient or almost sufficient to produce a solution of the full equations of motion, as we now explain.

Backgrounds which solve (\ref{unbroken}) admit Killing spinors.  Using Killing spinor identities \cite{Kallosh:1993wx}, one can show that
some of the equations of motion are automatically satisfied. The local supersymmetry of the action depending on bosons $\phi^i$ and on fermions $\psi^\alpha$ means that
\be
\delta_{\varepsilon} S(\phi^i, \psi^\alpha)= {\delta S\over \delta \phi^i} \delta_{\varepsilon} \phi^i+{\delta S\over \delta \psi^\alpha} \delta_{\varepsilon} \psi^\alpha=0~.
\label{KSI}\ee
Here, $\delta_{\varepsilon} \phi^i$ is the supersymmetry variation of the bosons and $\delta_{\varepsilon} \psi^\alpha$ is the supersymmetry variation of the fermions.
Unbroken supersymmetry of a bosonic configuration means that when the fermions vanish, $\psi^\alpha=0$,
\be
\delta_{\varepsilon} \psi^\alpha=X_{ A}^\alpha (\phi) \varepsilon ^A=0
\label{susy}\ee
and not all components of $\varepsilon ^A$ vanish.\footnote{Here $A$ runs over the number of supercharges in the N-extended supergravity; for us $N=2$.}
If we differentiate eq. (\ref{KSI}) over the fermions $\psi^{\beta}$ at $\psi^\alpha=0$, then using the facts that ${\delta S \over \delta \phi^i}$ and $\delta_{\varepsilon}\psi^\alpha$
only have terms which are either independent of the $\psi^\alpha$ or of quadratic or higher order in $\psi^\alpha$,  together with (\ref{susy}), we find
\be
 {\delta S\over \delta \phi^i} {\delta(\delta_{\epsilon} \phi^i)\over \delta \psi^\beta}=0~.
\label{KSI1}\ee
These identities require certain linear combination of the bosonic field equations to vanish. The variational derivative of the supersymmetry transformation of the bosons $\delta_{\varepsilon} \phi^i$ with respect to the fermions defines the relevant linear combinations of the equations of motion which are satisfied automatically, when the first order differential equations for unbroken supersymmetry
are solved.

Clearly, then, the simplification of the equations of motion in supersymmetric configurations depends both on details of the supersymmetric theory in question and on the
number of unbroken supersymmetries.  For instance, for minimal 5d supergravity, powerful techniques using Killing spinors were actually used to classify all supersymmetric solutions
in \cite{Gauntlett}.
For the particular case of 4D N=2 supergravity, the Killing spinor identities \cite{Kallosh:1993wx,ortinetal} were studed in detail in \cite{ortinetal,Hristov:2010eu}.
It was found there that if one has solved the first order differential equations for unbroken supersymmetry, in a background with a timelike Killing spinor,
the Einstein equations and the scalar equations of motion are automatically satisfied.
However, one still has to solve
equations of motion for the gauge fields \footnote{When the solution to the equations is given in terms of the gauge field  $A_\mu$, required in gauged supergravities, the Bianchi identities are automatically satisfied.}.
This is of course a considerable simplification.  Backgrounds with null Killing spinors are not quite as constrained; for details see \cite{ortinetal}.

In the next subsection, we go through the analysis of the SUSY variations for gauged N=2 supergravities.  We will focus on a class of metrics in D=4 spacetime $(t,x^m)$ ($m=1,2,3$) given by
\begin{align}\label{eq:metric}
  {\rm ds}^2 = e^{2U} ( {\rm d}t ^2+f_m dt \, dx^m) - e^{-2U}
  \gamma_{mn}{\rm d}x^m {\rm d}x^n\ ,
\end{align}
where $e^{2U}$ and $f_m$ and $\gamma_{mn}$ are independent of $t$.  This ansatz is sufficiently general to capture black holes as well as solutions with planar symmetry,
like domain walls and plane symmetric solutions of interest in applications of gauge/gravity duality to condensed matter problems.
Such metrics often admit  Killing vectors (and of course Killing spinors, in the case of BPS solutions).

\subsection{Solutions with unbroken supersymmetry}

For simplicity, we consider here the time-independent static metric in \eqref{eq:metric} with $f_m=0$ which corresponds to a frame
\be
e_t^0= e^U \qquad e_x^1= e^{-U} v_x^1 \qquad e_y^2= e^{-U} v_y^2 \qquad e_z^3= e^{-U} v_z^3
\ee
where $\gamma_{mn} dx^m dx^n= \delta _{ij} v^i v^j$.
We will now discuss how this frame is canonically determined in supersymmetric attractors.

If we have a supersymmetric attractor, then some Killing spinor  $\varepsilon_A(x^m)$ solves the
condition for unbroken supersymmetry arising from the gravitino supersymmetry variation, $\delta_\varepsilon \psi_{\mu A} = 0$, defined in
\eqref{susy-gravi}. Such spinors anti-commute, but we may also use
 commuting spinors (by expanding in a basis of Grassmann variables)  and relate them to Killing vectors.
 If by abuse of notation we also call the commuting spinor $\epsilon_A$, then we can define
 \be
 \bar \epsilon_A \equiv i (\epsilon^A)^\dagger \gamma_0 \qquad
{\rm
 and}\qquad
 V^{A}_{\mu B} \equiv i \bar \epsilon^{A} \gamma_{\mu} \epsilon_{B}~.
 \ee
  \label{vis}
 One can then show that this implies that $V^{A}_{\mu A}$ is a timelike or null Killing vector (see \cite{ortinetal} for proofs); we will pursue the timelike case below.
 The dependence of the Killing spinor on $x^m$ is
 \be
 \varepsilon_A(x^m)= e^{U}  \varepsilon_A^0
 \ee
 where $\varepsilon_A^0$ is a constant spinor.  Furthermore, we get a tetrad canonically determined by the Killing spinor as follows.  We set
 \be
 \label{one}
 V^\mu \partial_\mu = \sqrt{2} \partial_t \, , \qquad 
 \hat V^i = {1\over \sqrt{2}}V^{I}_{J \mu} \sigma^{i J}_{~~I} dx^{\mu}
\ee
 with $\sigma^i$ ($i=1,2,3$) the Pauli matrices.  The equations (\ref{one})  define a tetrad canonically determined by the Killing spinor $\epsilon_A$;
 rescaling each by $e^{-U}$ gives rise to an orthonormal tetrad (as in \S4\ of the first reference in \cite{ortinetal}).

 Before proceeding to solve the equations guaranteeing the existence of unbroken supersymmetry, we describe how the
 unbroken supersymmetry implies the equations of motion.  As in \cite{Hristov:2010eu}, denote by
 ${\cal E}^\mu_{a}$, ${\cal E}^{\mu}_{\Lambda},  {\cal E}_i,$ and ${\cal E}_u$ the equations of motion for the vielbein $e_\mu^a$, the gauge
 field $A_{\mu}^{\Lambda}$, and the scalars $z^i$ and $q^u$ (so e.g. ${\cal E}^\mu_a = 0$ are the Einstein equations).  Then the strategy explained
 above yields the following Killing spinor identities for supersymmetric backgrounds:
 \be
 {\cal E}_{\Lambda}^{\mu} i f_i^{\Lambda} \gamma_{\mu} \epsilon^A \epsilon_{AB} + {\cal E}_i \epsilon_B = 0
 \ee
 \be
 {\cal E}_a^\mu (-i\gamma^a \epsilon^A) + {\cal E}_{\Lambda}^{\mu} (2 \bar L^{\Lambda} \epsilon_B \epsilon^{AB}) = 0
 \ee
 \be
 {\cal E}_{u} {\cal U}^{u}_{\alpha A}\epsilon^A = 0~.
 \ee
  Here,
${\cal U}_{u}^{A\alpha}$ is related to the metric on the quaternionic K\"ahler moduli space by the equation
\begin{equation}
h_{uv} = {\cal U}_u^{A \alpha} {\cal U}_v^{B \beta} C_{\alpha \beta} \epsilon_{AB}
\end{equation}
with $C_{\alpha \beta} = - C_{\beta \alpha}$ the flat $Sp(2n_H)$-invariant metric (see \S5 of \cite{Andrianopoli:1996cm} for further discussion).

One can prove directly from these Killing spinor identities that if the gauge field equations of motion ${\cal E}_{\Lambda}^{\mu} = 0$ are satisfied,
all of the remaining field equations will be satisfied \cite{ortinetal,Hristov:2010eu}.\footnote{This is for a Killing spinor that yields a timelike Killing vector; in the null case, one must also
check one of the components of the Einstein equations explicitly \cite{ortinetal}.}
So apart from the BPS conditions which we discuss now, to find a
supersymmetric attractor point, one needs only to check the gauge field equations of motion (and the Bianchi identities).
As we already discussed, for the attractor points, these equations are purely algebraic.

{\it Gravitino}.  At the attractor point the equation
\be
\delta_\varepsilon \psi_{\mu A} = 0,
\ee
 simplifies significantly, and so does the integrability condition \cite{Ortin:2004ms}:
 \be
\delta_\varepsilon \psi_{\mu A}=0 \qquad \Rightarrow [D_a, D_b] \varepsilon_A= -{1\over 4} R_{ab}{}^{ cd}\gamma_{cd} \varepsilon_A= -(D_{[a }\tilde \delta\psi_{b]} + \tilde \delta\psi_a \tilde \delta\psi_b )_{AB} \varepsilon^B~
 \ee
 where the derivative $\partial _a$ in $D_{a }$ drops, and only the spin connection remains. Here, the fermionic shift is defined in (\ref{susy-gravitino}).
 $A_a$, $\omega_{a A}{}^B$, $T^-_{ab}$ and  $S_{A B}$ are constant at the attractor point. They are complicated functions of the vector multiplet scalars $z, \bar z$, the hypermultiplet scalars $q^u$, and the Killing vectors $k_\Lambda ^i, k_\Lambda ^u$
on the scalar manifolds,  and of the momentum maps $P_\Lambda ^x$ \cite{Andrianopoli:1996cm}.
The integrability condition for the Killing spinors is solved as follows.  It reads:
 \be
 \Big ({1\over 4} R_{ab}{}^{ cd}\gamma_{cd} - (\omega_{[a }\tilde \delta\psi_{b]} + \tilde \delta\psi_a \tilde \delta\psi_b )\Big )_{AB} \, \varepsilon^B=0~.
 \label{integr}\ee
To solve the algebraic equation (\ref{integr}) in a case with maximal unbroken supersymmetry, the terms in brackets must vanish. Otherwise, one finds some constraints on the Killing spinors, so that from the 2 4-component spinors $ \varepsilon_A$, only some fraction may be non-vanishing.  (When there are vanishing components of $\epsilon_A$,
then one should solve the subset of equations following from terms in brackets in (\ref{integr}) that multiply the non-vanishing components of the spinor). These equations
are algebraic conditions which define the values of $z, \bar z, q$ in terms of the other constant parameters in the action and  the anholonomy coefficients. We refer to \cite{Hristov:2010eu} for the details on gauged supergravity integrability equation for gravitino.

The universal feature of the integrability equations for the gravitino is that, at the attractor point, they relate the constant components  of the Riemann curvature to the constant values of scalars, vectors and parameters in the action.

{\it Gluino}. At the attractor point
\be
\delta_\varepsilon\lambda^{iA}= \tilde \delta \lambda ^{i AB} =0
\label{gaugino}\ee
where the fermionic shift $\tilde \delta \lambda$ is defined in (\ref{gluinoshift}). It is an an algebraic function of the scalars and the parameters appearing in the action and governing the moduli space geometry.
As in the gravitino case, one finds some constraints on $ \varepsilon_A$ and the remaining algebraic equations in (\ref{gaugino}) have to be satisfied.

{\it Hyperino}.  Here the situation is analogous:
at the attractor point
\be
\delta_\varepsilon\zeta_\alpha= \tilde \delta \zeta_\alpha =0
\label{hyperino}\ee
where  the fermionic shift are given in (\ref{susy-hyperino})
With account of the constraints on $ \varepsilon_A$,  the remaining algebraic equations in (\ref{hyperino}) have to be satisfied. They relate the attractor values of scalars to parameters in the action and the moduli space geometry.

We have seen in this section that to solve all equations of motion in a BPS configuration, one is required only to solve the gauge field equations explicitly;
the other equations are then implied by the Killing spinor identities.  However, it is often more convenient to find the attractor values of the scalars by minimizing
an attractor potential, whose form we now describe.  This has the advantage that the extrema of the attractor potential also determine the non-supersymmetric attractor points.

\section{Attractor Potential, BPS and Non-BPS  Critical Points}

In supergravity, generic equations of motion for scalars have terms with two spacetime derivatives, one spacetime derivative, and no spacetime derivatives:
\be
X^{\mu\nu} \partial_\mu \partial_\nu \phi + X^\mu \partial_\mu  \phi + {\partial   {\cal V}[\phi,...]\over \partial \phi} =0~.
\ee
Here the potential ${\cal V}[\phi,...]$ depends on scalar fields without derivatives, and it may depend on other spacetime dependent fields, like $F_{\mu\nu}$ etc.

The BPS solutions have some unbroken supersymmetry, and they solve scalar equations with at most first derivatives:
\be
Y^\mu \partial_\mu  \phi + {\partial   {\cal V}_{\rm BPS}[\phi,...]\over \partial \phi} =0~.
\ee
Here again, ${\cal V}_{\rm BPS}$ depends on scalars without derivatives and on some other spacetime dependent fields.

At attractor points the scalars are constant, so we are looking at the solutions of equations of motion with constant scalars. There is a potential for such supergravity attractors such that the variation of the scalars is an extremum of this attractor potential:
\be
 {\partial   {\cal V}_{\rm attractor}[\phi]\over \partial \phi} =0~.
\ee
In the case of $D=4$ $N=2$ gauged supergravity, the attractor potential is given by
\be
{\cal V}_{\rm attractor}(z, \bar z, q)= g^2 \Big( V(z, \bar z, q)-
(g_{i\bar
\jmath} { k}^i_{\Lambda}
k^{\bar \jmath}_\Sigma  + h_{uv}   { k}^u_{\Lambda}
k^{v}_\Sigma)
A^{\Lambda}_{a} A ^{a \Sigma}
\Big)
-\frac{1}{2}\epsilon^{abcd}
F_{ab}^{\Lambda}G_{cd \Sigma} ~.
\label{calV}\ee
Here $V(z, \bar z, q)$ is the standard potential given in eq. (\ref{pot2}), and $G_{cd \Sigma} $ also depend on scalars as shown in eq. (\ref{eq:defg}). $F$ and $G$ here are the solutions of the gauge field equations and Bianchi identities where $G$ is a functional of vectors and scalars.

The attractor potential has a beautiful form in terms of the fermionic shifts, defined in Sec. 3. It generalizes the  Ward identity  to vacua with non-vanishing constant vector potentials and field strengths. Namely as in eq. (9.47) of \cite{Andrianopoli:1996cm}, for N-extended supergravity vacua the scalar potential takes the form:
\be
\delta_A{}^B V(z, \bar z, q)= Z_{\alpha \beta} \delta_A\chi^\alpha \delta^B \bar \chi^\beta - 3 {\cal M}_{AC} \bar {\cal M}^{CB}
\ee
where $Z_{\alpha \beta}$ is the scalar dependent metric on moduli space, ${\cal M}_{AC}$ is the scalar dependent gravitino mass matrix, and $ \delta_A\chi^\alpha$ are the fermionic shifts.  Here $A=1,...,N$.

The potential governing our generalized attractor points for general gauged $N=2$ supergravities has a similar structure in terms of fermionic shifts, but the shifts $\tilde \delta_A  \chi^\alpha$ and the gravitino mass $\tilde {\cal M}_{AC}$ include terms depending on constant gauge fields and constant gauge field strengths:
\be
\delta_A{}^B {\cal V}_{\rm attractor} (z, \bar z, q)= Z_{\alpha \beta} \tilde \delta_A  \chi^\alpha \tilde \delta^B \bar \chi^\beta - 3 \tilde {\cal M}_{AC} \tilde {\bar{\cal M}}^{CB} ~.
\ee
When we differentiate the attractor potential with respect to the scalars, the derivative remains proportional to the fermionic shifts, since the potential is quadratic in them. Thus,  $ {\partial   {\cal V}_{\rm attractor}[\phi]\over \partial \phi} =0$ will be a consequence of the vanishing of the fermionic shifts, and hence of the presence of some unbroken supersymmetry.
In such cases, we have supersymmetric attractor points.

Of course, there are also non-supersymmetric attractors; these show up when $ {\partial   {\cal V}_{\rm attractor}[\phi]\over \partial \phi} =0$ but the fermionic shifts do not vanish.
In the case of black holes these have been studied in \cite{Ferrara:1996dd,Trivedi,Ceresole} and many subsequent works.

\section{Examples}

In this section, we discuss a few simple examples of our general construction.  For general supersymmetric solutions, as discussed in \S4, all equations of motion
are guaranteed to be satisfied if we sit at a minimum of the attractor potential and solve the gauge field equations of motion.  The novel ingredient is the central role played
by metrics of constant anholonomy in the preferred tetrad.  Therefore, here we just focus on solving the equations for constant $c_{ab}^{~~c}$ in an orthonormal tetrad; each such metric
should arise in a wide variety of gauged supergravities, and such metrics should be the universal geometries governing attractor points.  We will not specify
special K\"ahler and quaternionic K\"ahler manifolds and particular gaugings which, together with these metrics, give a full solution of supergravity; we expect the
metrics we find below to appear in many different specific examples.

We emphasize that all of the geometries we discuss below as examples of constant anholonomy have arisen before in other contexts.   Our main purpose here is just to exhibit them emerging, in a unified way, as the simplest and most generic solutions to the equations governing generalized attractor points.

\subsection{ AdS$_4$ and dS$_4$ solutions}

AdS$_4$ solutions of gauged N=2 supergravity are common and well studied. They are also a particular case of the example discussed below in \S6.2\ and in Appendix A:
AdS$_4$ is the special case of a Lifshitz geometry with $z=u=1$.

In contrast, dS solutions are not easy to find in gauged supergravities. Generically with non-compact gaugings one can find dS vacua, but typically, there are tachyonic
fluctuations in the spectrum and these critical points are unstable. Stable dS vacua in gauged N=2 supergravity were constructed in \cite{Fre:2002pd}. The ingredients include non-Abelian non-compact gauging, de Roo-Wagemans  rotation angles, and Fayet-Iliopoulos terms.

In terms of geometry, dS$_4$ spaces give one of the simplest examples of constant anholonomy and constant tangent-space curvature. The metric is
\be
ds^2 =  dt^2  - e^{2Ht }  d\vec x^2 ~.
\ee
The constant  anholonomy coefficients are (no summation in tangent indices $\hat k$)
\be
c_{0 \hat k}^{\hat k}=-  c_{\hat k 0}^{\hat k }= -H~.
\ee
 The non-zero Riemann curvature components in the coordinate basis are (no summation on curved space indices  $i$ or $j$)
 \be
 R^{t}_{~i,i,t}= R^{i }_{~j,j,t} = - R^{i}_{~j,i,j}=
 - e^{2Ht} H^2\, ,
\qquad
 R^{i}_{~t,i,t}=
 -  H^2~.
 \ee
 Some of them are time-dependent.  The tangent space Riemann curvature components are constant. All non-vanishing components of it are
\be
R_{abcd}= \pm H^2~.
\ee
\subsection{Lifshitz solutions}

In light of gauge/gravity duality, in order to find gravity duals of field theories with spatial translation and time translation symmetries, and (in the special case $\gamma_{xx}=\gamma_{yy}$ below) spatial
rotation symmetry, it is useful to study the metric ansatz:
\be
ds^2 = e^{2U(r)} dt^2  - e^{-2U(r)}  \left(   \gamma_{xx}^2(r)dx^2 + \gamma_{yy}^2(r)dy^2 + dr^2\right)~.
\ee
A  vielbein  and the associated dual tetrad are given by:
\bea
e^{0}_t & = & e^{U}, \qquad e^1_x  =  e^{-U}\gamma_{xx} , \qquad  e^2_y = e^{-U}\gamma_{yy} , \qquad e^{3}_r = e^{-U} \, \nonumber \\
\tilde e_{0}& = & e^{-U}\partial_t, \qquad \tilde e_1  =  {e^{U} \over \gamma_{xx}} \partial_x, \qquad \tilde e_2 = {e^{U} \over \gamma_{yy}} \partial_y, \qquad \tilde e_{3} = e^{U} \partial_r \, .
\eea
The non-trivial anholonomy coefficients are
\be
c_{30}^{~~0} = -U^\prime e^{U}
\ee
and
\be
c_{31}^{~~1} = \left(U^\prime - {\gamma_{xx}^\prime \over \gamma_{xx}}\right) e^{U}, \, \,\,\,
c_{32}^{~~2} = \left(U^\prime - {\gamma_{yy}^\prime \over \gamma_{yy}}\right) e^{U}~.
\ee
It is easy to find the generic solution with constant anholonomy; the solution of the first equation gives
\be
e^U = b_1 r + b_2~.
\ee
Without loss of generality, we can set $b_2=0$ by shifting $r$.

The last two equations $c_{31}^{~~1} = {\rm const} \equiv b_{3x}$ and $c_{32}^{~~2} = {\rm const} \equiv b_{3y}$ now become:
\bea
{{\rm d\over dr}} {\rm log}~\gamma_{xx} & = & (b_1 -  b_{3x}) e^{-U},\nonumber \\
{{\rm d\over dr}} {\rm log}~\gamma_{yy} & = & (b_1 -  b_{3y}) e^{-U}~.
\eea
This yields
\bea
\gamma_{xx}(r) & = & b_{4x} r^{(b_1 - b_{3x}) \over b_{1}}~, \nonumber \\
\gamma_{yy}(r) & = & b_{4y} r^{(b_1 - b_{3y}) \over b_{1}}~.
\eea

The upshot is that the generic attractor point with planar symmetry is a Lifshitz solution \cite{KLM} (with, in general, different scaling in the $t,x,y$ directions).  
The supersymmetry of a class of  Lifshitz solutions to some
particular N=2 gauged supergravities
was established in \cite{Cassani,Halmagyi:2011xh}.  Constant scalars, vectors, and field strengths support the solutions in \cite{Cassani,Halmagyi:2011xh}, and give nice
examples of non-trivial solutions of the algebraic attractor equations of gauged N=2 supergravity described in this paper.  The components of the Riemann curvature for the Lifshitz
geometry are given in Appendix A; the reader can verify that in the tangent space, $R^{a}_{~bcd}$ is constant.

In general, as in \cite{KLM}, we expect that for fully consistent solutions (which do not require e.g. imaginary fluxes to support the background), there will
be constraints on the scaling dimensions (e.g. the $z \geq 1$ constraint in \cite{KLM}).
Some familiar solutions arise for particular values of the parameters in the solutions above: $b_{3x}=b_{3y} = 0$ corresponds to AdS$_4$ solutions, while solutions with
$b_1 = b_{3x}=b_{3y}$ correspond to AdS$_2$$ \times$ R$^2$ geometries.  That is, these geometries are the extreme limits of Lifshitz solutions where the
dynamical critical exponent $z$ goes to $1$ or $\infty$.

\subsection{Schr\"odinger  solutions}

The Schr\"odinger metric takes the form
\be
ds^2 = r^{2z} du^2 - 2r^2 du dv-  - {dr^2\over r^2} - r^2 dx^2~.
\ee
A convenient vielbein $e_{\mu}^a dx^\mu$ is given by
\be
e^0 = {r^z\over 2} du
\, , \qquad
e^1 = r^z du - 2r^{2-z} dv
\, , \qquad
e^2 = r dx
\, , \qquad
e^3 = {dr\over r}~.
\ee
It is a simple matter to compute the $e^{\nu}_{a} = \eta^{ab}e^{b}_{\mu}g^{\mu\nu}$, and from these the anholonomy coefficients.
One finds:
\be
c_{30}^{~~0} =- z\, , \qquad c_{30}^{~~1} = 4 (1 - z)\, , \qquad c_{31}^{~~1} = 2 - z\, , \qquad c_{32}^{~~2} = -1~.
\ee
So for all values of $z$, the Schr\"odinger geometries are also examples of spaces of constant anholonomy, and can be expected to
arise at generalized attractor points in various N=2 gauged supergravities.  The components of the Riemann curvature tensor for the
Schr\"odinger geometry are given in Appendix B, and are constants in the tangent space basis, as expected.
One specific gauged supergravity admitting Schr\"odinger solutions
is discussed in \cite{Halmagyi:2011xh}.  The paper \cite{Perlmutter} finds a general correspondence between AdS and
Schr\"odinger solutions;  interestingly, their Schr\"odinger vacua sit at the same attractor values of the moduli as the related AdS vacua.

\section{Discussion}
Many supersymmetric and non-supersymmetric  attractor solutions of supergravity are known in D=4.
The solutions with constant scalars fit into the general structure of supergravity attractor points described above.   An interesting new class of
supersymmetric and plane-symmetric solutions with constant scalars and vectors has been identified recently in \cite{Cassani,Halmagyi:2011xh}. These are non-relativistic solutions of N=2 gauged supergravity with constant scalars, including Lifshitz \cite{KLM},
Schr\"odinger \cite{Son,McGreevy}, and AdS solutions. The solutions in \cite{Cassani,Halmagyi:2011xh} were based on the study of particular N=2 supergravities.  They give specific examples of generalized attractors in gauged N=2 D=4 supergravity.   One should expect to find many further examples of interesting attractor points, governed by the algebraic equations discussed here.

We have focused in this paper on the solutions where the scalars are constant over the entire space-time.
A more general class of solutions (with or without unbroken supersymmetry) interpolates between two different vacua, each of which is
characterized by fixed scalars and constant (tangent space) vector fields.  The starting and ending points of such interpolating solutions are described by the universal algebraic equations presented in this paper.  In all cases, supersymmetric or non-supersymmetric, the scalars at the attractor points are constant,
the geometry is characterized by constant anholonomy coefficients, and the equations of motion are algebraic.  The extremization of the gauged supergravity attractor potential provides the equations which fix the constant values of scalars.
For supersymmetric solutions, one requires the vanishing of the fermionic shifts defined in \S3.   More general extrema of the gauged supergravity attractor potential provide non-supersymmetric critical points of gauged supergravity.

It would be interesting to explore analogous simplifications which occur at attractor points in extended supergravities in D $\neq$ 4; to give explicit examples of
supersymmetric attractors in N=2 theories with simple canonical choices of the special K\"ahler and quaternionic K\"ahler manifold; and, especially, to give a complete
classification of the metrics that can possibly arise at attractor points.  For instance, under certain assumptions, a classification of near-horizon geometries of extremal
black holes was given in \cite{reall}; having such a classification for extremal black brane geometries would be valuable, and would constitute a gauge/gravity duality
``prediction'' for the possible ground states of holographic matter.
In addition, finding in full generality the allowed flows between different kinds of attractor
points would be quite interesting, as it would shed light on possible renormalization group flows between dual field theories with different kinds of scale invariance.
We leave these problems for the future.

Finally, the attentive reader will have noticed that the reduction of the equations of motion to algebraic equations occurs in this paper due to the simplicity of working
with fields that are constant in the tangent space, and is not tied to supersymmetry.  For this reason, we suspect it should be possible to find analogous algebraic
solutions in many theories even without supersymmetry - in much the same way that the original attractor mechanism for extremal black holes has been generalized to non-supersymmetric
theories as well \cite{Trivedi,Ceresole}.

\section*{Acknowledgments}
We are grateful to Sandip Trivedi for numerous stimulating conversations about the attractor mechanism, black hole and black brane solutions, and possible attempts at
classifications thereof.  We also thank D. Garfinkle, G. Horowitz, A. Linde, R. Myers,  S. Shenker and E. Silverstein for enjoyable discussions.  We
thank M. Mulligan and S. Yaida for helpful comments on a draft.  S.K. acknowledges the hospitality of the Aspen Center for Physics while he was thinking about related issues.

\section{Appendix A: Lifshitz geometry}

Here, we work with the anisotropic  Lifshitz geometry
\be
ds^2 = r^{2z}dt^2  - r^{2u} dx^2 - r^2 dy^2 - {dr^2\over r^2} ~.
\ee
(Solutions with different scaling in the $t,x,y$ directions were considered previously in e.g. \cite{Pal}.)
The vielbeins are
\be
e^0_t= r^z ,  \qquad e^1_x= r^u , \qquad e^2_y= r , \qquad e^3_r= {1\over r}
\label{e}\ee
The anholonomies are
\be
c_{03}{}^0= {z}\, ,  \quad c_{13}{}^1= {u}\, , \quad c_{23}{}^2= {1}\, , \quad c_{30}{}^0=- {z}\, , \quad c_{31}{}^1= -{u}\, , \quad c_{32}{}^2= -{1} \, .
\ee
In the coordinate basis $t,x,y,r$, we find the following non-zero components of the Riemann curvature:
\be
\nonumber
R^{t}_{~x,x,t} = r^{2u} u z\, , \qquad R^{t}_{~y,y,t} = r^2 z\, , \qquad R^{t}_{r,r,t} = {z^2 \over r^2}~,
\ee
\be
\nonumber
R^{x}_{~t,x,t} = r^{2z}u z\, , \qquad R^{x}_{~y,y,x} = u r^2\, , \qquad R^{x}_{~r,r,x} = {u^2\over r^2}~,
\ee
\be
\nonumber
R^{y}_{~t,y,t} = r^{2z} z\, , \qquad R^{y}_{~x,y,x} = -u r^{2u}\, , \qquad R^{y}_{~r,r,y} = {1\over r^2}~,
\ee
\be
\nonumber
R^{r}_{~t,r,t} = r^{2z}z^2\, , \qquad R^{r}_{~x,r,x} = -u^2 r^{2u}\, , \qquad R^{r}_{~y,r,y} = -r^2.
\ee
In the tangent space, with $0,1,2,3$ indices now corresponding to the tetrad in (\ref{e}), we find instead
\be
\nonumber
R^{0}_{~1,1,0} =  uz\, ,  \qquad R^{0}_{~2,2,0} =  z\, ,  \qquad R^{0}_{3,3,0} = z^2\, ,  \qquad
\ee
\be
\nonumber
R^{1}_{~0,1,0} = - uz\, ,  \qquad R^{1}_{~2,2,1} = -u\, ,  \qquad R^{1}_{~3,3,1} = -u^2\, ,  \qquad
\ee
\be
\nonumber
R^{2}_{~0,2,0} = -z\, ,  \qquad R^{2}_{~1,2,1} = 1\, ,  \qquad R^{2}_{~3,3,2} = -1\, ,  \qquad
\ee
\be
\nonumber
R^{3}_{~0,3,0} = -z^2\, ,  \qquad R^{3}_{~1,3,1} = u^2\, ,  \qquad R^{3}_{~2,3,2} = 1\, .
\ee
Happily, these are constant for all values of $z$ and $u$.

\section{Appendix B: Schr\"odinger curvatures}

Here, we work with the Schr\"odinger spacetime described in \S6.3.  The non-zero Riemann curvature components in the coordinate basis
$u,v,x,r$ are given by:
\be
\nonumber
R^{u}_{~u,v,u} = r^2\, , \qquad R^{u}_{~x,x,u} = r^2\, , \qquad R^{u}_{~r,r,u} = {1\over r^2}~,
\ee
\be
\nonumber
R^{v}_{~u,v,u} = r^{2z}\, , \qquad R^{v}_{~v,v,u} = -r^2\, , \qquad R^{v}_{~x,x,u} = -r^{2z} (-1 + z)~,
\ee
\be
\nonumber
R^{v}_{~x,x,v} = r^2,~R^{v}_{~r,r,u} = -2 r^{-4 + 2z} (-1+z) z,~R^{v}_{~r,r,v} = {1\over r^2}~,
\ee
\be
\nonumber
R^{x}_{~u,x,u} = z r^{2z}\, , \qquad R^{x}_{~r,r,x} = {1\over r^2}\, , \qquad R^{r}_{~u,r,u} =r^{2z}(1 + (-1 + z) z)~,
\ee
\be
\nonumber
R^{r}_{~u,r,v} = -r^2\, , \qquad R^{r}_{~v,r,u} = -r^2\, , \qquad R^{r}_{~x,r,x} = -r^2~.
\ee

In the tangent space, with now $0,1,2,3$ indices corresponding to $e_0, e_1, e_2, e_3$ in \S6.2, we instead find:
\be
\nonumber
R^{0}_{~1,0,0} = -2\, , \qquad R^{0}_{~1,1,0} = 1\, , \qquad R^{0}_{~2,2,0} = 4(-1 + z)~,
\ee
\be
\nonumber
R^{1}_{~2,2,1} = 1\, , \qquad R^{0}_{~3,3,0} = 8 (-1+z) z\, , \qquad R^{0}_{~3,3,1} = 1~,
\ee
\be
\nonumber
R^{1}_{0,0,0} = 2\, , \qquad R^{1}_{~0,1,0} = -1\, , \qquad R^{1}_{~2,2,0} = 1~,
\ee
\be
\nonumber
R^{1}_{~3,3,0} = 1\, , \qquad R^{2}_{~0,2,0} = 4-4z\, , \qquad R^{2}_{~0,2,1} = -1~,
\ee
\be
\nonumber
R^{2}_{~1,2,0} = -1\, , \qquad R^{2}_{~3,3,2} = -1\, , \qquad R^{3}_{~0,3,0} = -8 (-1+z) z,
\ee
\be
\nonumber
R^{3}_{~0,3,1} = -1\, , \qquad R^{3}_{~1,3,0} = -1\, , \qquad R^{3}_{~2,3,2} = 1~.
\ee
We see that happily, the Schr\"odinger spacetime is characterized by constant Riemann tensor (in the tangent space) for arbitrary values of $z$.

\end{document}